\documentclass[12pt]{article}
\begin{document}
\title{Semileptonic decay of $B_{c}$ meson into S wave charmonium in a QCD potential model with coulombic part as perturbation.}
\author{ $^{1}$Krishna Kingkar Pathak and $^{2}$ D K Choudhury \\
$^{1}$Deptt. of Physics,Arya Vidyapeeth College,Guwahati-781016,India\\
e-mail:kkingkar@gmail.com\\
$^{2}$Deptt. of Physics, Gauhati University, Guwahati-781014,India}
\date{}
\maketitle
\begin{abstract}
We present the semileptonic decay of $B_{c}$ meson in a QCD potential model with the coulombic part of the Cornell potential $-\frac{4\alpha _{S}}{3r}+br+c$ as perturbation. Computing the slope and curvature of Isgur Wise function in this approach, we study the pseudoscalar and vector form factors for the transition of $B_{c}$ meson to its S wave charmonium $c\overline{c}$ states. Numerical estimates of widths for the transitions of $B_{c}\rightarrow J/\psi(\eta_{c} )l\nu_{l}$ are presented. The results are found to be in good agreement with other theoretical values. \\

\end{abstract}
\section{Introduction}
The investigation of weak decays of mesons composed of a heavy quark and antiquark gives a very important insight in the heavy quark dynamics.The exclusive semileptonic decay processes of heavy mesons generated a great excitement not only in extracting the most accurate values of Cabbibo-Kobayashi Maskawa(CKM) matrix elements but also in testing diverse theoretical approaches to describe the internal structure of hadrons. The great virtue of semileptonic decay processes is that the effects of the strong interaction can be separated from the effects of the weak interaction into a set of Lorentz-invariant form factors, i.e., the essential informations of the strongly interacting quark/gluon structure inside hadrons. Thus, the theoretical problem associated with analyzing semileptonic decay processes is essentially that of calculating the weak form factors.

 The decay properties of the$B_{c}$ meson are of special interest, since it is the only heavy meson consisting of two heavy quarks with different flavor. This difference of quark flavors forbids annihilation into gluons. As a result, the excited $B_{c}$ meson states lying below the $B_{D}$ meson threshold undergo pionic or radiative transitions to the pseudoscalar ground state which is considerably more stable than corresponding charmonium or bottomonium states and decays only weakly. The CDF Collaboration reported the discovery of the $B_{c}$ ground state in $p\overline{p}$ collisions already more than ten years ago [1]. However, up till recently its mass was known with a very large error. Now it is measured with a good precision in the decay channel $B_{c}\rightarrow{J/\psi\pi}$.  More experimental data on masses and decays of the Bc mesons are expected to come in near future from the Tevatron at Fermilab and the Large Hadron Collider (LHC) at CERN. The estimates of the Bc decay rates indicate that the c quark transitions give the dominant contribution while the b quark transitions and weak annihilation contribute less. However, from the experimental point of view the $B_{c}$ decays to charmonium are easier to identify. Indeed, CDF and D0 observed the $B_{c}$ meson and measured its mass analyzing its semileptonic and nonleptonic decays $B_{c}\rightarrow{J/\psi l\nu}$.\\
There are many theoretical approaches to the calculation of exclusive $B_{c}$ semileptonic decay modes. Some of them are: QCD sum rules [ 2, 3, 4], the relativistic quark model [5, 6, 7] based on an effective Lagrangian
describing the coupling of hadrons to their constituent
quarks, the quasipotential approach to the relativistic
quark model [8, 9, 10], the instantaneous nonrelativistic
approach to the Bethe-Salpeter(BS) equation [11], the
relativistic quark model based on the BS equation [12,
13], the QCD relativistic potential model [14], the relativistic quark-meson model [15], the nonrelativistic quark
model [16], the covariant light-front quark model [17],
and the constituent quark model [18, 19, 20, 21] using BSW(Bauer, Stech, and Wirbel) model [22] and
ISGW(Isgur, Scora, Grinstein, and Wise) model [23].
The purpose of this paper is to extend a QCD potential model[24] with coulombic part as perturbation to calculate
the hadronic form factors and decay widths for the exclusive semileptonic decay of $B_{c}$ meson.\\
Recently, we have reported the slope and curvature of I-W function for $D$ and $B$ mesons with the coulombic part of the potential as the perturbation in two particular renormalisation schemes $\overline{MS}$ and $V$ scheme [24]. Instead of using a particular renormalisation schemes, here in this manuscript we use the strong coupling constant as a scale dependent parameter and compute the slope and curvature of I-W funtion for $B_{c}$ meson with a different set of mass input parameters than that of ref.[24]. We then use the I-W function to study the form factors and decay rates of $B_{c}$ meson into its $S$ wave charmonium $c\overline{c}$ states within the framework of the potential model.\\

The rest of the paper is organised as follows : section 2 contains the formalism with its subsections containing the model wavefunction, Masses, hadronic form factors and decay widths. In section 3  we place our results and conclusions.\\ 
  
\section{Formalism}
\subsection{The wavefunction}
The wavefunction computed by Dalgarno method [25,26] with coulombic part $-\frac{4\alpha_s}{3r}+c$ of the potential as perturbation and linear part $br$ as parent has been reported in ref.[24]. For completeness we summarise the main equations in this section. However the alternate approach of choosing the linear part as perturbation has been reported earlier [27,28].\\  
The total wave function corrected upto first order with normalisation is [24]
\begin{eqnarray}
\psi_{coul}\left(r\right)&=&\psi^{\left(0\right)}\left(r\right)+\psi^{\left(1\right)}\left(r\right)\\&=&\frac{N_{1}}{2\sqrt \pi}\left[\frac{Ai(\left(2\mu b\right)^{\frac{1}{3}}+\rho_{01})}{r}-\frac{4\alpha_{s}}{3}\left(\frac{a_{0}}{r}+a_{1}+a_{2}r\right)\right]
\end{eqnarray}

where $ N_{1}$ is the normalisation constant for the total wave function $\psi_{coul}\left(r\right)$ with subscript 'coul' means coulombic potential as perturbation.\\
where $\rho_{0n}$ are given as \cite{29,30}: 
\begin{equation}
\rho_{0n}=-[\frac{3\pi\left(4n-1\right)}{8}]^{\frac{2}{3}}
\end{equation}
\begin{equation}
a_{0}=\frac{0.8808 \left(b\mu\right)^{\frac{1}{3}}}{\left(E-c\right)}-\frac{a_{2}}{\mu\left(E-c\right)}+\frac{4 W^{1}\times 0.21005}{3\alpha_{s}\left(E-c\right)}
\end{equation}
\begin{equation}
 a_{1}=\frac{ba_{0}}{\left(E-c\right)}+\frac{4\times W^{1}\times0.8808\times \left(b\mu\right)^{\frac{1}{3}}}{3\alpha_{s}\left(E-c\right)}-\frac{0.6535\times \left(b\mu\right)^{\frac{2}{3}}}{\left(E-c\right)}
\end{equation}
\begin{equation}
a_{2}=\frac{4\mu W^{1}\times0.1183}{3\alpha_{s}}
\end{equation}
\begin{equation}
W^{(1)}=\int_{0}^{+\infty}r^{2} H^{\prime}\left|\psi^{(0)}\left(r\right)\right|^{2} dr
\end{equation}
and
\begin{equation}
E= -\left(\frac{b^{2}}{2\mu}\right)^{\frac{1}{3}} \rho_{0n}
\end{equation}
here $b$ and $c$ are the model input parameters as is used in our previous works [24,27,28]. $'n'$ is the principal quantum no.(n=1 for ground state),$\mu$ is the reduced mass of mesons and $\alpha_{s}$ is the strong running coupling constant.

\subsection{Estimation of meson masses }
 The energy shift of mass splitting due to spin interaction in the perturbation theory reads[31]
 \begin{equation}
\Delta E=\int \Psi^{*}\left(\frac{32\pi\alpha_{s}}{9}\delta^{3}\left(r\right)\frac{S_{i}.S_{j}}{m_{i}.m_{j}}\right)\psi d^{3}r
\end{equation}
leading to
\begin{equation}
\Delta E=\frac{32\pi\alpha_{s}}{9m_{i}m_{j}}\left(S_{i}.S_{j}\right)|\psi\left(0\right)|^{2}
\end{equation}
Taking the energy shift into account, meson mass is expressed as
\begin{equation} 
M_{p}=m_{i}+m_{j}-\frac{ 8\pi\alpha_{s}}{3m_{i}m_{j}} |\psi\left(0\right)|^{2}
\end{equation}
\begin{equation} 
M_{v}=m_{i}+m_{j}+\frac{ 8\pi\alpha_{s}}{9m_{i}m_{j}} |\psi\left(0\right)|^{2}
\end{equation}

since $(S_{i}.S_{j})=-3/4$ for pseudoscalar meson and $(S_{i}.S_{j})=1/4$ for vector meson . \\
In the model approach, the wave function at the origin $ |\psi\left(0\right)|$ develops singularity. Therefore
for computing $ |\psi\left(0\right)|^{2}$ we use the phenomenological law given by Igi and Ono[32] as
\begin {equation}
|\psi\left(0\right)|^{2}=k{\mu^{1.52}}
\end{equation} 
here k is a proportionality constant.
In eq.(1), the strong coupling constant connected to the potential is a function of the momentum as
\begin{equation}
\alpha_{s}\left(\mu_{1}^{2}\right)=\frac{4\pi}{\left(11-\frac{2n_{f}}{3}\right)ln\left(\frac{\mu_{1}^{2}}{\Lambda^{2}}\right)}
\end{equation}
where $n_{f}$ is the number of flavour and $\mu_{1}$ is the renormlistion scale related to the constituent quark mass and $\Lambda$ is the QCD scale which is taken as 0.150 GeV by fixing $\alpha_{s}=0.118$ at the Z boson mass(91 GeV). We use the most common renormalisation scale as $\mu_{1}=4\frac{m_{i} m_{j}}{m_{i}+m_{j}}$ with $n_{f}=4$ [32,33] and then evaluate  $\alpha_{s}$ for $B_{c}$, $\eta_{c}$ and $J/\psi$ mesons.
 
For numerical calculation,we use the model input parameters as, $m_{c}=1.55$ GeV and $m_{b}=4.79$GeV  and $k=0.05$ to compute the masses of $B_{c}$, $\eta_{c}$ and $J/\psi$ mesons and are shown in table 1.  
\begin{table}[h]
\begin{center}
\caption{Masses of $c\overline{b}$ and $c\overline{c}$ state }
\vspace {.2in}
\begin{tabular}{c c c c}\hline
mesons&$|\psi\left(0\right)|^{2}$ in $GeV^{3}$ & {values of $\alpha_{s}$ }& $M_{p,v}$ in GeV\\\hline 

$B_{c}$&0.061 &0.22 &6.274[our work ] \\

 & & &6.277 Expt\cite{PDG2010}\\ 
 & & &6.270 \cite{8}\\
 & & &6.302 \cite{PRD.56.4133}\\\hline

$\eta_{c}$&0.034 &0.25 &3.070[our work ] \\

 & & &2.980 Expt\cite{PDG2010}\\
 & & &2.979 \cite{8}\\
 & & &3.088 \cite{PRD.74.014012}\\\hline
$J/\psi$&0.034 &0.25 &3.109 [our work]\\
 & & &3.097 Expt\cite{PDG2010}\\
 & & &3.096 \cite{8} \\
 & & &3.168 \cite{PRD.74.014012}\\\hline

 \end{tabular}
\end{center}
\end{table}

\subsection{Pseudoscalar and Vector form factors}

In the case of the final $c\bar c$ states corresponds to the $J=0$, $\eta_c$ states, as the matrix element of any axial current $A^\mu$ between the two pseudoscalar mesons vanishes, only vector current $V^\mu$ contributes. Unlike in the case of electromagnetic current of the charged pions, here the  vector current $V^\mu=\bar c\gamma^\mu b$ is not conserved as  $q_\mu V^\mu\propto (m_b-m_c)\neq 0$. So the matrix element of the hadronic current, $V^\mu$  between the two $J^{P}=0^{-}$ mesons is expressed in terms of two form factors $f_\pm (q^{2})$ as

\begin{equation}\label{eq.p}
\langle\eta_c(p^{'})|V^\mu|B_c(p)\rangle=f_+(q^2)(p+p^{'})_\mu+f_-(q^2)(p-p^{'})_\mu
\end{equation}
Where $q=p-p^{'}=k_1+k_2$ is the four momentum transfer and $f_+(q^2)$ and $f_-(q^2)$ are the dimensionless weak transition form factors corresponds to $B_c\rightarrow\eta_c$, which are functions of the invariant $q^2$. Here $q^2$ varies within the range $m_\ell^2\leq q^2\leq(m_{B_c}-m_{\eta_c})^2=q_{max}^2$.\\

The transition between the pseudoscalar $B_c$ and the vector $J/\psi(p^{'},\epsilon)$ mesons depends on four independant form factors as,
\begin{eqnarray}\label{eq.v}
\langle J/\psi(p^{'},\epsilon)|\bar c\gamma^\mu b|B_c(p)\rangle&=&2i\epsilon{\mu\nu\alpha\beta}\frac{\epsilon_\nu p^{'}_\alpha p_\beta}{M_{B_c}+M_{J/\psi}}V(q^2)\\
\langle J/\psi(p^{'},\epsilon)|\bar c\gamma^\mu\gamma_5b|B_c(p)\rangle&=&(M_{B_c}+M_{J/\psi})\left[\epsilon^\mu-\frac{\epsilon\cdot qq^\mu}{q^2}\right]A_1(q^2)\nonumber\\
&&-\epsilon\cdot q\left[\frac{(p+p^{'})^\mu}{M_{B_c}+M_{J/\psi}}-\frac{(M_{B_c}-M_{J/\psi})q^\mu}{q^2}\right]A_2(q^2)\nonumber\\
&&2M_{J/\psi}\frac{\epsilon\cdot q q^\mu}{q^2}A_0(q^2)
\end{eqnarray}
In the present study we treat $B_c$ system similar to $D$ $(c\bar d$) system as in ref.[37] and extend HQET to express  Eqns. \ref{eq.p} and \ref{eq.v} as  
\begin{equation}
\frac{1}{\sqrt{M_{B_c}M_{\eta_c}}}\langle\eta_c(v^{'})|V^\mu|B_c(v)\rangle=(v+v^{'})^\mu\xi(\omega)
\end{equation}
\begin{equation}
\frac{1}{\sqrt{M_{B_c}M_{J/\psi}}}\langle J/\psi(v^{'},\epsilon_3)|V^\mu|B_c(v)\rangle=i\epsilon^{\mu\nu\alpha\beta}\epsilon_\nu v^{'}_\alpha v_\beta\xi(\omega)
\end{equation}
\begin{equation}
\frac{1}{\sqrt{M_{B_c}M_{J/\psi}}}\langle J/\psi(v^{'},\epsilon_3)|A^\mu|B_c(v)\rangle=[(1+\omega)\epsilon^\mu-(\epsilon\cdot v)v^{'\mu}]\xi(\omega)
\end{equation}
Where $\xi(\omega)$ is the Isgur Wise function.For small, nonzero recoil, it is conventional to write the Isgur-Wise function as \cite{38,39} :
\begin{eqnarray}
\xi\left(v.v^{\prime}\right)\nonumber&=&\xi(\omega)\\&=&1-\rho^{2}\left(\omega-1\right)+ C\left(\omega-1\right)^{2}+...
\end{eqnarray}
 where $\omega$ is given by,
\begin{equation}
w=v\cdot v^{'}=\frac{\left[m_{B_c}^2+m_{c\bar c}^2-q^2\right]}{2m_{B_c}m_{c\bar c}}
\end{equation}

The quantity $\rho^{2}$  is the slope of I-W function at $\omega=1$ and known as charge radius :\\

\begin{equation}
\rho^{2}= \left. \frac{\partial \xi}{\partial \omega}\right.|_{\omega=1}
\end{equation}
The second order derivative is the curvature of the I-W function known as convexity parameter :\\

\begin{equation}
C=\left .\frac{1}{2}\right. \left(\frac{\partial^2 \xi}{\partial \omega^{2}}\right)|_{\omega=1}
\end{equation}
For the heavy-light flavor mesons the I-W function can also be written as \cite{40,27} :\\

\begin{equation}
\xi\left(\omega \right)=\int_{0}^{+\infty} 4\pi r^{2}\left|\psi\left(r\right)\right|^{2}\cos pr dr
\end{equation}
where

\begin{equation}
p^{2}=2\mu^{2}\left(\omega-1\right)
\end{equation}
Using eq.1,eq.21 and eq.25, we compute the slope and curvature of I-W function as\\
  $\rho_{B_{c}}^{2}=0.67 $ and $C_{B_{c}}=0.06$.\\

Consequently, the form factors $f_\pm(q^2)$ correspond to the $c\bar c(\eta_c)$ final state are related to the Isgur Wise function as \cite{37,41}
\begin{equation}\label{fpm}
f_\pm(q^2)=\xi(\omega)\frac{m_{B_c}\pm m_{\eta_c}}{2\sqrt{m_{B_c}m_{\eta_c}}}
\end{equation}
And those related to the $J/\psi$ as the final hadronic state are given by 
\begin{equation}\label{eq.vformfactora}
V(q^2)=A_2(q^2)=A_0(q^2)=\left[1-\frac{q^2}{(M_{B_c}+M_{J/\psi})^2}\right]^{-1}A_1(q^2)=\frac{(M_{B_c}+M_{J/\psi})^2}{4M_{B_c}M_{J\psi}}\xi(\omega)
\end{equation}
It is evident from Eqn. \ref{eq.vformfactora} that at $q^2\rightarrow0$
\begin{equation}
V(q^2)=A_2(q^2)=A_1(q^2)=A_0(q^2)
\end{equation}
Thus knowing the masses and Isgur Wise function of the transition $B_c\rightarrow c\bar c(\eta_c,J/\psi)\ell^+\nu_\ell$, we will be able to compute  respective form factors correspond to $B_c\rightarrow c\bar c(\eta_c,J/\psi)\ell^+\nu_\ell$ transitions.

It should be noted that by virtue of transversality of the lepton current $l_{\mu} = l\gamma_{\mu}(1 +\gamma_{5})\nu{l}$ in the limit $m_{l}\rightarrow{0}$, the probabilities of semileptonic decays into $e^{+}\nu_{e}$ and $\mu^{+}\nu_{\mu}$ are independent of $f_{-}$ . Thus, in calculation of these particular decay
modes of $B_{c}$ meson this form factor can be consistently neglected [3].\\

The differential semileptonic decay rates can be expressed in terms of
these form factors by \\
(a) $B_{c}\to Pe\nu$ decay  ($P=\eta_{c}$)
\begin{equation}
  \label{eq:dgp}
  \frac{{\rm d}\Gamma}{{\rm d}q^2}(B_{c}\to Pe\nu)=\frac{G_F^2 
  \Delta^3 |V_{qb}|^2}{24\pi^3} |f_+(q^2)|^2.
\end{equation}
(b) $B_{c}\to Ve\nu$ decay ($V=J/\psi$)
The decays rate in transversally and longitudinally polarized vector
mesons are defined by[42]
\begin{equation}
  \label{eq:dgl}
\frac{{\rm d}\Gamma_L}{{\rm d}q^2}=\frac{G_F^2
\Delta|V_{qb}|^2}{96\pi^3}\frac{q^2}{M_{B}^2}
|H_0(q^2)|^2,  
\end{equation}
\begin{equation}
  \label{eq:dgt}
\frac{{\rm d}\Gamma_T}{{\rm d}q^2}=
\frac{{\rm d}\Gamma_+}{{\rm d}q^2}+\frac{{\rm d}\Gamma_-}{{\rm d}q^2}
=\frac{G_F^2\Delta|V_{qb}|^2}{96\pi^3}\frac{q^2}{M_{B}^2}
\left(|H_+(q^2)|^2+|H_-(q^2)|^2\right). 
\end{equation}
where helicity amplitudes are given by the following expressions
\begin{equation}
  \label{eq:helamp}
  H_\pm(q^2)=\frac{2M_{B_{c}}\Delta}{M_{B_{c}}+M_V}\left[V(q^2)\mp
\frac{(M_{B_{c}}+M_V)^2}{2M_{B_{c}}\Delta}A_1(q^2)\right],
\end{equation}
\begin{equation}
  \label{eq:h0a}
  H_0(q^2)=\frac1{2M_V\sqrt{q^2}}\left[(M_{B_{c}}+M_V)
(M_{B_{c}}^2-M_V^2-q^2)A_1(q^2)-\frac{4M_{B}^2\Delta^2}{M_{B_{c}}
+M_V}A_2(q^2)\right].
\end{equation}
Thus the total decay rate is given by[42]
\begin{equation}
  \label{eq:dgv}
\frac{{\rm d}\Gamma}{{\rm d}q^2}(B_{c}\to Ve\nu)=\frac{G_F^2
\Delta|V_{qb}|^2}{96\pi^3}\frac{q^2}{M_{B_{c}}^2}
\left(|H_+(q^2)|^2+|H_-(q^2)|^2   
+|H_0(q^2)|^2\right),
\end{equation}
where $G_F$ is the Fermi constant, $V_{qb}$ is CKM matrix element ($q=c$),
\[\Delta\equiv|{\bf\Delta}|=\sqrt{\frac{(M_{B_{c}}^2+M_{P,V}^2-q^2)^2}
{4M_{B_{c}}^2}-M_{P,V}^2}.
\]
Integration over $q^2$ of these formulas gives the total rate
of the corresponding semileptonic decay. The computed results for the form factors and the total decay rate is shown in table 2,table 3 and table 4 respectively and compared with the other theoretical values.
\begin{table}[htbp]
\begin{center}
\caption{Form factors for $B_c\rightarrow\eta_c\ell^+\nu_\ell$}\label{table:fpm}
\begin{tabular}{ccccc}
\hline
$work$&$f_+(0)$&$f_-(0)$&$f_+(q_{max}^2)$&$f_-(q_{max}^2)$\\\hline
our work &0.886&-0.304&1.065&-0.365\\

Vinodkumar[37]  &	0.71&	-0.25&	0.78&	-0.28\\
Faustov[42 ] &0.47 & &1.07 &\\
Choi and Ji[43]&0.546 & &1.035 &\\
\hline
\end{tabular}
\end{center}
\end{table}
\begin{table}[htbp]
\begin{center}
\caption{Form factors for $B_c\rightarrow J/\psi\ell^+\nu_\ell$}\label{table:vfm}
\begin{tabular}{cccc}
\hline
$work$&$V(0)$&$A_1(q_{max}^2)$&$V(q_{max}^2)$\\
\hline
Our work &0.891&	0.941&	1.0622\\
Vinod kumar[37] &0.70 &0.75 &0.77\\
Faustov[42]&0.49 &0.88 &1.34 \\
\hline
\end{tabular}
\end{center}
\end{table}
\begin{table}[htbp]
\begin{center}
\caption{Decay width for $B_c\rightarrow c\bar c(\ell^+\nu_\ell)$ in $10^{-15}GeV$}\label{decay}
\begin{tabular}{ccc}
\hline
&$B_c\rightarrow\eta_c(\ell^+\nu_\ell)$&$B_c\rightarrow J/\psi(\ell^+\nu_\ell)$\\
\hline
  &  16.4&    39.8\\\hline

\cite{6}&10.7&28.2\\
\cite{9}&5.9&17.7\\
\cite{PRD.65.014017}&14.2&34.4\\
\cite{PRD.62.014019}&11.1&30.2\\
\cite{PRD.56.4133}&8.31&20.3\\
\cite{NPB.569.473}&$11\pm1$&$28\pm5$\\
\cite{PRD.61.034012}&2.1&21.6\\
\cite{YF.62.1868}&8.6&17.2\\
\cite{NPB.440.251}&10&42\\
\hline
\end{tabular}
\end{center}
\end{table}

\section{Results and Discussion}
In this paper, we have computed the slope and curvature of Isgur Wise function for $B_{c}$ meson, considering the coulombic part of the cornell potential as perturbation. Presumably, the values of $\rho^{2}$ and $C$ are found to be smaller than the other theoretical values for $D$ and $B$ mesons. The reason is supposed to be due to the truncating of higher terms in the Airy function(which is an infinite series). To avoid the singularity in the wavefunction at the origin, we also use a model independent phenomenological parametrisation with the same mass input parameters to calculate the masses of  $c\overline{b}$ and $c\overline{c}$ states in the ground state and is shown in table.1. These masses are in good agreement with the experimental result.\\
The computed I-W function and masses are used to study the hadronic form factors using HQET in analogy to $D$ meson system and is compared with other available results which are shown in table.2 and table.3. The decay widths for $B_c\rightarrow c\bar c(\ell^+\nu_\ell)$ and $B_c\rightarrow J/\psi(\ell^+\nu_\ell)$ are shown in table.4 and found to be in  agreement with other results.\\

\end{document}